\documentclass[3p,times]{elsarticle}

\usepackage{ecrc}

\volume{00}

\firstpage{1}

\journalname{Physica A}

\runauth{P.~Jizba and J.~Korbel}

\jnltitlelogo{Physica A}

\usepackage{amssymb}
\usepackage{amsthm}

\usepackage[usenames,dvipsnames]{color}
\usepackage{amsmath}
\usepackage{dsfont}
\usepackage{graphicx}
\usepackage{algorithm}
\usepackage{float}
\usepackage{algpseudocode}
\usepackage{multirow}
\newfloat{algorithm}{t}{lop}

\definecolor{darkred}{rgb}{.8,0,0}

\definecolor{darkblue}{rgb}{0,0,.7}

\begin{document}

\begin{frontmatter}



\dochead{}
%% e.g. \dochead{17th International Conference on Dynamical Processes in Excited States of
%% Solids}

%%%%%%%%%%%%%%%%%%%%%%%%%%%%%%%%%%%%%%%%%%%%%%%%%%%%%%%%%%%%%%%%%%%%%%%%%%%%%%%%%%%%%%%%%%%%%%%%%%%%%%%%%%%%%%%%%%%%%%%%%%%%%%%%%%%%%%%%%%%%%%
\title{Comment on ``Generalized Shannon--Khinchin axioms and uniqueness theorem for pseudo-additive entropies'' [Physica A 411 (2014) 138]}
%%%%%%%%%%%%%%%%%%%%%%%%%%%%%%%%%%%%%%%%%%%%%%%%%%%%%%%%%%%%%%%%%%%%%%%%%%%%%%%%%%%%%%%%%%%%%%%%%%%%%%%%%%%%%%%%%%%%%%%%%%%%%%%%%%%%%%%%%%%%%%


\author[FNSPE,FU]{Petr Jizba}
\ead{p.jizba@fjfi.cvut.cz}

\author[ZJU,FNSPE]{Jan Korbel}
\ead{korbeja2@fjfi.cvut.cz}

\address[FNSPE]{Faculty of Nuclear Sciences and Physical Engineering, Czech Technical University in Prague, B\v{r}ehov\'{a} 7,
11519, Prague, Czech Republic}
\address[FU]{Institute of Theoretical Physics, Freie Universit\"{a}t in Berlin, Arnimallee 14, 14195 Berlin, Germany}
\address[ZJU]{Department of Physics, Zhejiang University, Hangzhou 310027, P. R. China}

\vspace{-6mm}
\begin{abstract}
Recently in [Physica A 411 (2014) 138] Ili\'{c} and Stankovi\'{c} have suggested that there
may be problem for the class of hybrid entropies introduced in  [P.~Jizba and T.~Arimitsu, Physica A 340 (2004) 110]. In this Comment we point out that the problem can be traced down to the $q$-additive entropic chain rule and to a peculiar behavior of the DeFinetti--Kolmogorov relation for escort distributions. However, despite this, one can still safely use the proposed hybrid entropies in most of the statistical-thermodynamics considerations.
\end{abstract}


\end{frontmatter}

%%%%%%%%%%%%%%%%%%%%%%%%%%%%%%%%%%%%%%%%%%%%%%%%%%%%
\section{Introduction}
%%%%%%%%%%%%%%%%%%%%%%%%%%%%%%%%%%%%%%%%%%%%%%%%%%%%

In their recent paper~\cite{is},  Ili\'{c} and Stankovi\'{c} have suggested (in Remark~5.3)
that there may be problem for the class of hybrid entropies introduced in~\cite{ja} and elaborated in our recent paper~\cite{jk}. In particular, they argued that the entropies in question do not
satisfy the fourth axiom in generalized Shannon--Khinchin axioms (J-A axioms) introduced in~\cite{ja}. The fourth axiom in question is basically the $q$-additive entropic chain rule. They do so by generalizing  the single-parameter J-A axiomatics by inserting yet another independent parameter $\alpha$ into axioms. The new parameter appears in the definition of the conditional entropy where it specifies the order of the weighting escort distribution.  Ili\'{c} and Stankovi\'{c} have succeeded in solving their new axiomatic system and found that ensuing solutions split into 4 distinct classes: i) $q=1, \alpha = 1$, ii) $q\neq 1, \alpha =1$, iii) $q=1, \alpha \neq 1$ and
iv) $q\neq 1, \alpha \neq 1$. Interestingly enough, when $q = \alpha$ [i.e. special case in iv)]
then the axioms turn out to be identical with the J-A axioms but their solution does not coincide with our hybrid entropy but rather with the usual Tsallis entropy. This latter finding is particularly intriguing because Tsallis entropy is not usually affiliated with the Kolmogorov--Nagumo means (which explicitly enter the J-A axiomatics and Ili\'{c}--Stankovi\'{c} calculations).

In this comment, we are not concerned with this, the most interesting part of
Ili\'{c}--Stankovi\'{c} fine paper. Rather, we seek to clarify
their remarks about the r\^{o}le of our hybrid entropy in their axiomatic system.
Instead of following the route outlined in Ili\'{c}--Stankovi\'{c} paper, we simply consider
the hybrid entropy as given in~\cite{ja,jk} and try to see where the points of incompatibility
with the J-A axiomatics arise. We do so within the framework of escort distributions,
which offer a particularly illuminating tool for this task.
The upshot of this analysis is that our hybrid entropies satisfy the Tsallis-type  $q$-additivity condition for any two independent events.
As long as the events (or systems) considered are dependent, the $q$-additive entropic chain rule is not satisfied. We trace down the root cause of this behavior to a peculiar behavior of the  De~Finetti--Kolmogorov relation for escort distributions.

%%%%%%%%%%%%%%%%%%%%%%%%%%%%%%%%%%%%%%%%%%%%%%%%%%%%%%%%%%%%%%%%%%%%%%%%%%%%%%%%%%%%%%%%%%%%%%%%%%%%%%%%
\section{Hybrid entropy and generalized Shannon--Khinchin axioms}
%%%%%%%%%%%%%%%%%%%%%%%%%%%%%%%%%%%%%%%%%%%%%%%%%%%%%%%%%%%%%%%%%%%%%%%%%%%%%%%%%%%%%%%%%%%%%%%%%%%%%%%

We recall first the so-called J-A axioms for hybrid entropy~\cite{ja} , namely
\begin{enumerate}
\item \emph{continuity}: $\mathcal{D}_q(p)$ is a continuous function of all arguments
\item \emph{maximality}: for given $n$ is $\mathcal{D}_q(p)$ maximal for $p = \left({1}/{n}, \dots, {1}/{n}\right)$
\item \emph{expansibility}: $\mathcal{D}_q(p_1,\dots,p_n,0)= \mathcal{D}_q(p_1,\dots,p_n)$
\item \emph{J-A additivity}: $\mathcal{D}_q(A,B) \ = \ \mathcal{D}_q(A) + \mathcal{D}_q(B|A)\ + \ (1-q) \mathcal{D}_q(A) \mathcal{D}_q(B|A)$, where
\begin{eqnarray*}
\mathcal{D}_q(B|A) \ = \ f_q^{-1} \left( \sum_k P(q)_k f_q (\mathcal{D}_q(B|A = A_k)) \right),
\end{eqnarray*}
and $f_q$ is some positive-definite invertible function on $\mathds{R}^+$.
\end{enumerate}
The J-A axioms 1-4 were introduced in~\cite{ja} as a unifying one-parameter framework for both Tsallis and R\'{e}nyi entropy.
As such, the axioms appear quite instructive because they allow to address the currently popular concept of
$q$-nonextensive entropies from an entirely new angle of view.  Of course, this is true provided some non-trivial entropy functional $\mathcal{D}_q$  satisfying these axioms exists. In Ref.~\cite{jk} it was argued that the unique
solution of the J-A axioms has the form.
\begin{equation}
\mathcal{D}_q(p) \ = \ \frac{1}{1-q} \left(e^{-(1-q) \sum_k P(q)_k \ln p_k} -1\right).
\label{hybrid}
\end{equation}
Here $P(q)_k= p_k^q/\sum_l p_l^q$ is the escort distribution of $q$-th order. Before we point out
the potential problem with this form, let us discuss first the respective J-A axioms in the
connection with entropy functional~(\ref{hybrid}). This is a worthy pursuit because a)
it allows to asses the extend of damage caused and  b) it suggests potential rectifications to be taken.
In addition, MaxEnt distributions derived from $\mathcal{D}_q$ have a very rich practical applicability (see,e.g., Ref.~\cite{jk,valluri})
and it would be nice to retains some of the desired properties of $\mathcal{D}_q$ even in a limited sense.

Let us now take a closer look at  $\mathcal{D}_q$ in the context of the J-A axiomatic .\\[2mm]
%
%%%%%%%%%%%%%%%%%%%%%%%%%%%%%%
\noindent{\em{1) continuity:}}
%%%%%%%%%%%%%%%%%%%%%%%%%%%%%%
Apparently, the entropy $\mathcal{D}_q(p)$ is continuous in all arguments for arbitrary $n$.\\[2mm]
%
%%%%%%%%%%%%%%%%%%%%%%%%%%%%%%
\noindent{\em{2) maximality:}}
%%%%%%%%%%%%%%%%%%%%%%%%%%%%%%
Maximality axiom has been extensively discussed in Ref.~\cite{jk} and it has been shown that hybrid entropy obeys the maximality axiom for $q \geq \frac{1}{2}$.\\[2mm]
%
%%%%%%%%%%%%%%%%%%%%%%%%%%%%%%%%%
\noindent{\em{3) expansibility:}}
%%%%%%%%%%%%%%%%%%%%%%%%%%%%%%%%%
There is no doubt that $\mathcal{D}_q(p)$ is expansible function. This is clear from the
sum structure and the fact that  $p_k^q \cdot \ln p_k = 0$ if $p_k =0$ and $q>0$.\\[2mm]
%
%%%%%%%%%%%%%%%%%%%%%%%%%%%%%%%%%%%%%%%%%%%%%%%%%%%%%%%%%%
\noindent{\em{4a) additivity rule -- independent events:}}
%%%%%%%%%%%%%%%%%%%%%%%%%%%%%%%%%%%%%%%%%%%%%%%%%%%%%%%%%%

First, Let us discuss two independent events $A$ and $B$ with respective distributions $p = \{p_k\}$ and $q=\{q_k\}$ and
associated {\em escort distributions}
\begin{eqnarray}
&&P(q)_{k} \ = \ \frac{p_k^q}{\sum_i p_i^q} \, \leftrightarrow \, p_k \ = \ \frac{[P(q)_k]^{1/q}}{\sum_i [P(q)_i]^{1/q}}\, ,\nonumber \\
&&Q(q)_{k} \ = \ \frac{q_k^q}{\sum_i q_i^q} \, \leftrightarrow \, q_k \ = \ \frac{[Q(q)_k]^{1/q}}{\sum_i [Q(q)_i]^{1/q}}\, ,
\end{eqnarray}
We also introduce a function $f_q(x)$ as
\begin{eqnarray}
f_q(x) \ = \ \ln \exp_q (x) \ = \ \ln [1+(1-q)x]^{1/(1-q)}   \ \; \Rightarrow \; \ f_q^{-1}(x) \ = \  \ln_q \exp (x)\ = \ \frac{1}{(1-q)}\left[e^{(1-q)x} -1\right].
\end{eqnarray}
The function $f_q(x)$ coincides with the Kolmogorov--Nagumo function from 4th axiom as found in~\cite{ja}.

The relevance of  $\exp_q$ and $\ln_q$ stems from their close connection to $q$-calculus~\cite{bor}, and from the fact that
they are precisely identical with $q$-exponential and $q$-logarithmic functions used in Tsallis statistics~\cite{yam}.
In particular, if we define the $q$-addition as
\begin{equation}
a \oplus_q b \ = \ a + b + (1-q)ab\, ,
\end{equation}
then it is easy to check that $f_q$ fulfills the relation
\begin{equation}\label{fq}
f_q(a\oplus_q b) \ = \ f_q(a)+f_q(b)
\, .
\end{equation}
For $q = 1$, the $q$-addition $\oplus_q$ reduces to a standard addition operation and $f_q$ boils down to an identity function.

With the help of the relation~\eqref{fq}, the 4th axiom can be equivalently expressed as
\begin{equation}
f_q(\mathcal{D}_q(A,B)) \ = \ f_q(\mathcal{D}_q(A)) + f_q(\mathcal{D}_q(B|A))\, .
\label{additivity}
\end{equation}
Expression $f_q(\mathcal{D}_q(p))$ is also known as the Acz\'{e}l--Dar\'{o}czy (additive) entropy~\cite{ad}.

By employing the above $f_q$-mapping we will simplify number of cumbersome mathematical steps and
stay comparatively close to the reasonings used in the Ili\'{c}--Stankovi\'{c} article~\cite{is}.
Under the $f_q$-mapping the respective Acz\'{e}l--Dar\'{o}czy entropies read
\begin{eqnarray}
f_q({\mathcal{D}}_q(A)) &=&  - \frac{\sum_j p_j^q \ln p_j}{\sum_j p_j^q} \
   = \ - \frac{1}{q} \sum_k P(q)_k \ln P(q)_k  + \ln\sum_k P(q)^{1/q}_k \nonumber\\[1mm]
   &=& \frac{1}{q} {\mathcal{S}}(P(q))  - \frac{(1-q)}{q}{\mathcal{I}}_{1/q}(P(q))\, ,\label{3}\\[2mm]
f_q({\mathcal{D}}_q(B)) &=&  - \frac{\sum_j q_j^q \ln q_j}{\sum_j q_j^q} \ = \
  %\frac{1}{1-q} \left(e^{-\frac{(1-q)}{q} \sum_k Q(q)_k \ln Q(q)_k  + (1-q) \sum_k Q(q)^{1/q}} -1\right)\nonumber\\[1mm]
%   &=&
\frac{1}{q} {\mathcal{S}}(Q(q))  - \frac{(1-q)}{q}{\mathcal{I}}_{1/q}(Q(q))\, .
   \label{4}
\end{eqnarray}
Here ${\mathcal{S}}$ and ${\mathcal{I}}_{1/q}$ denote  Shannon's and R\'{e}nyi's entropy (of the order $1/q$), respectively.

Applying J-A additivity [in the form (\ref{additivity})] alongside with the entropy form (\ref{hybrid}) and the fact that for {\em independent} events $\mathcal{D}_q(B|A = A_k) = \mathcal{D}_q(B)$  [which implies that $\mathcal{D}_q(B|A)$ from ``4'' is simply $\mathcal{D}_q(B)$] we can write
\begin{eqnarray}
  f_q(\mathcal{D}_q(A,B)) \! &=& \!   -\frac{\sum_j p_j^q \ln p_j}{\sum_j p_j^q} -
  \frac{\sum_k q_k^q \ln q_k}{\sum_k q_k^q} \nonumber \\[2mm]
  &=& \!
 \frac{1}{q} [{\mathcal{S}}(Q(q)) + {\mathcal{S}}(P(q)) ]  - \frac{(1-q)}{q}[{\mathcal{I}}_{1/q}(Q(q)) + {\mathcal{I}}_{1/q}(P(q))]\nonumber \\[2mm]
 &=& \! \frac{1}{q} {\mathcal{S}}(Q(q)P(q))  - \frac{(1-q)}{q}{\mathcal{I}}_{1/q}(Q(q)P(q)).
\label{eq:1}
\end{eqnarray}
On the last line we have used the additivity of both Shannon and R\'{e}nyi entropy. By realizing that
\begin{eqnarray}
Q(q)_kP(q)_l \ = \ \frac{q_k^q p_l^q }{\sum_{kl} q_k^q p_l^q } \ \equiv \ \frac{r_{kl}^q}{\sum_{kl} r_{kl}^q}\, ,
\end{eqnarray}
we can write the last line of (\ref{eq:1}) as [cf. also (\ref{3})  and (\ref{4})] as
\begin{eqnarray}
\frac{1}{q} {\mathcal{S}}(Q(q)P(q))  - \frac{(1-q)}{q}{\mathcal{I}}_{1/q}(Q(q)P(q)) \ = \
-\frac{\sum_{kl} r_{kl}^q \ln r_{kl}}{\sum_{kl} r_{kl}^q}\, .
\end{eqnarray}
This is indeed nothing but $f_q(\mathcal{D}_q(A,B))$. So we see that the hybrid entropy  $\mathcal{D}_q$ does satisfy the J-A additivity axiom (basically $q$-additivity) for {\em independent} events.
Note also that the validity of the J-A additivity rule for the hybrid entropy is in this case a direct consequence of the additivity of both Shannon and R\'{e}nyi entropy.\\[2mm]
%
%%%%%%%%%%%%%%%%%%%%%%%%%%%%%%%%%%%%%%%%%%%%%%%%%%%%%%%%%%
\noindent{\em{4b) additivity rule -- dependent events:}}
%%%%%%%%%%%%%%%%%%%%%%%%%%%%%%%%%%%%%%%%%%%%%%%%%%%%%%%%%%%
according to 4th axiom, we can calculate $f_q(\mathcal{D}_q(B|A))$ in two ways:
\begin{eqnarray}
f_q(\mathcal{D}_q(B|A)) &=& f_q(\mathcal{D}_q(A,B)) - f_q(\mathcal{D}_q(A))\, ,\label{eq: 3}\\
  f_q(\mathcal{D}_q(B|A)) &=& \sum_k P(q)_k f_q (\mathcal{D}_q(B|A = A_k))\, .\label{eq: 2}
\end{eqnarray}
 We denote the conditional probability
$p(B=B_k|A=A_l) = r_{k|l}$ so that $r_{kl} = p(B=B_k,A=A_l) = p_l r_{k|l}$. With this
we can rewrite (\ref{eq: 3}) as
\begin{eqnarray}
f_q(\mathcal{D}_q(B|A)) \ &=& \ \frac{\sum_{j} p_{j}^q \ln
p_{j}}{\sum_{k} p_{k}^q} - \frac{\sum_{ij}
r_{ij}^q \ln r_{ij}}{\sum_{kl} r_{kl}^q}  \nonumber \\[2mm]
&=& \ \frac{1}{q} \left[{\mathcal{S}}(R(q))
- {\mathcal{S}}(P(q))\right]  - \frac{(1-q)}{q}\left[{\mathcal{I}}_{1/q}(R(q))
- {\mathcal{I}}_{1/q}(P(q))\right].
%  \frac{1}{1-q} \left[\exp\left(\frac{(1-q)}{q} {\mathcal{S}}() +
% \frac{(q-1)^2}{q} {\mathcal{I}}_{1/q}(P(q)) \right)
% -1\right]
\label{13}
\end{eqnarray}
Here we have employed the notation $R(q)_{ij} = r_{ij}^q/\sum_{kl}
r_{kl}^q$ or equivalently $r_{ij} =
[R(q)_{ij}]^{1/q}/\sum_{kl}[R(q)_{kl}]^{1/q}$. We can bring (\ref{13}) into more suggestive form by employing
the well known chain rules for Shannon and R\'{e}nyi entropies, namely
\begin{eqnarray}
&&{\mathcal{S}}(R(q)) \ = \ {\mathcal{S}}(P(q)) \ + \
{\mathcal{S}}([R(q)/P(q)])\, ,\nonumber \\[2mm]
&& {\mathcal{I}}_{1/q}(R(q)) \ = \ {\mathcal{I}}_{1/q}(P(q)) \ + \
{\mathcal{I}}_{1/q}([R(q)/P(q)])\, ,
\end{eqnarray}
where the conditional probability $[R(q)/P(q)]_{i|j}$ is defined via the usual De~Finetti--Kolmogorov relation $R(q)_{ij}  = P(q)_j \ \! [R(q)/P(q)]_{i|j}$.
%$r_{i|j}^q/\sum_{i} r_{i|j}^q$,
%denotes the conditional escort distribution
With this  the last identity in (\ref{13}) can be written succinctly as
\begin{eqnarray}
f_q(\mathcal{D}_q(B|A)) \ = \ \frac{1}{q} {\mathcal{S}}([R(q)/P(q)])  -
\frac{(1-q)}{q}{\mathcal{I}}_{1/q}([R(q)/P(q)])\, .
\label{13a}
\end{eqnarray}
In view of (\ref{3})-(\ref{4}) this seems to be genuinely correct and trouble-free form of the conditional hybrid entropy $\mathcal{D}_q(B|A)$. This is even bolstered by the fact that both ${\mathcal{S}}$ and ${\mathcal{I}}_{1/q}$ are well defined for conditional distributions.  However, there are two (closely connected) problems with the formula (\ref{13a}).

First problem resides in the fact that $R(q)_{ij}$ in the above employed De~Finetti--Kolmogorov relation  $R(q)_{ij}  = P(q)_j \ \! [R(q)/P(q)]_{i|j}$ cannot represent a joint probability distribution even though it derives from the genuine joint distribution $r_{ij}$.
%This can be seed directly from the fact that it does not
%
%First, $\mathcal{D}_q(B|A)$ is supposed to refer to conditioning with respect to original (non-escort) distributions, while (\ref{13a}) is %formulated in terms of escort distributions.
This is clear because the marginal distribution $P(q)_j$ cannot be obtained from $R(q)_{ij}$ by summing over index $i$.
In this connection we should recall obvious but underappreciated fact about the De~Finetti--Kolmogorov relation, namely that
$r_{kl} \ = \ p_l r_{k|l} \nLeftrightarrow R(q)_{kl}  = P(q)_l \ \!
[R(q)/P(q)]_{k|l}$  for $[R(q)/P(q)]_{k|l} = r_{k|l}^q/ \sum_n r_{n|l}^q$.
This follows from the simple chain of reasonings:
\begin{eqnarray}
r_{kl} \ = \ p_l r_{k|l} \, \Leftrightarrow \,  r_{kl}^q \ = \ p_l^q r_{k|l}^q
\, \Leftrightarrow \, \frac{r_{kl}^q }{\sum_{mn} r_{mn}^q} \ = \ \frac{p_l^q
r_{k|l}^q}{\sum_{mn} r_{mn}^q} \ = \ \frac{p_l^q
r_{k|l}^q}{\sum_{n} (p_n^q \sum_{m} r_{m|n}^q) }  \ \neq \  \frac{p_l^q}{\sum_{n}
p_n^q}
\frac{r_{k|l}^q}{\sum_{m} r_{m|l}^q } \ \equiv \ \tilde{R}(q)_{kl}\,  .
\end{eqnarray}
Here $\tilde{R}(q)_{kl}$ denotes the correct joint distribution. The equality $R(q)_{kl} = \tilde{R}(q)_{kl}$ holds only when $\sum_{mn} r_{mn}^q = \sum_{mn} p_n^q r_{m|l}^q$, for all indices ``$l$''.
%which is also the condition for $R(q)_{kl}$ being a joint distribution.
By subtracting two $\sum_{mn} r_{mn}^q$ with different $l$'s on the right-hand side (say $l$ and $l'$) we get
\begin{eqnarray}
0 \ = \ \sum_m \left(r_{m|l}^q - r_{m|l'}^q\right)\,\,\,\, \mbox{for all} \,\,\, l,l'\, .
\end{eqnarray}
The latter can be satisfied only when $r_{m|l}$ is constant for all $l$'s, i.e., when the two events involved are independent. This in turns says that $R(q)_{ij}$ represents a genuine joint distribution only in the case of independent events.

%Clearly, one can make even stronger statement by saying that the De~Finetti--Kolmogorov
%relations for statistics with escort distributions of the different order $q$ are not equivalent.

Second, should we used the defining relation (\ref{eq: 2}) for the conditional hybrid entropy  together with the fact that
\begin{eqnarray}
f_q(\mathcal{D}_q(B|A = A_k)) \ = \  - \frac{\sum_l r_{l|k}^q \ln r_{l|k}}{\sum_m r_{m|k}^q}\, ,
\end{eqnarray}
we would obtain
\begin{eqnarray}
f_q(\mathcal{D}_q(B|A)) \ = \  - \sum_k P(q)_k \frac{\sum_l r_{l|k}^q \ln r_{l|k}}{\sum_m r_{m|k}^q}
\label{12a}
\end{eqnarray}
In terms of escort distributions the latter  can be recast into the form
\begin{eqnarray}
f_q(\mathcal{D}_q(B|A)) \ &=& \ \frac{1}{q} \left[\tilde{{\mathcal{S}}}(R(q))
- {\mathcal{S}}(P(q))\right]  - \frac{(1-q)}{q}\left[{\mathcal{I}}_{1/q}(R(q))
- {\mathcal{I}}_{1/q}(P(q))\right].
%
%
%
% \frac{1}{1-q} \left\{\exp\left[\frac{(1-q)}{q} \left[{\mathcal{S}}(R(q))
%- {\mathcal{S}}(P(q))\right] \ \right. \right. \nonumber \\[2mm]
%&-&  \left. \left. \ \frac{(1-q)^2}{q}\left(\sum_{k} P(q)_{k} \ln \sum_{n,m} P(q)_{n}^{1/q} [R(q)/P(q)]_{m|k}^{1/q}
% - {\mathcal{I}}_{1/q}(P(q))\right)\right] -1\right\}.
\label{12aa}
\end{eqnarray}
Where $\tilde{{\mathcal{S}}}(R(q)) = - \sum_{kl} \tilde{R}(q)_{kl} \ln R(q)_{kl}$.
Note, in particular, that (\ref{12aa}) differs from (\ref{13}) only in the $\tilde{{\mathcal{S}}}(R(q))$ term. The equivalence holds if and only if $\tilde{{\mathcal{S}}}(R(q))  = {{\mathcal{S}}}(R(q))$. In order to better understand when this is satisfied we write
\begin{eqnarray}
\tilde{{\mathcal{S}}}(R(q)) - {{\mathcal{S}}}(R(q))  \ = \ - \sum_{kl} R(q)_{kl} \left( \frac{\tilde{R}(q)_{kl}}{R(q)_{kl}} -1\right) \ln R(q)_{kl} \, ,
\end{eqnarray}
and use the relation
\begin{eqnarray}
\frac{\tilde{R}(q)_{kl}}{R(q)_{kl}} \ = \ \frac{\sum_{m,n} P(q)_m \ \! r_{n|m}^q}{\sum_{n} r_{n|k}^q } \ = \ \frac{\sum_{n} \langle r_{n|\bullet}^q \rangle}{\sum_{n} r_{n|k}^q }
% \ \leq \
%\frac{\sum_{n} \max_l r_{n|l}^q }{\sum_{n} r_{n|k}^q }
\, .
\label{20ab}
\end{eqnarray}
Here we may observe that even when $R(q)_{kl}$ is zero for some pair $\{k,l\}$ the fraction ${\tilde{R}(q)_{kl}}/{R(q)_{kl}}$ is not singular.
With the help of the Min-Max theorem for means~\cite{Hardy1959} and  Eq.~(\ref{20ab}) we can write
\begin{eqnarray}
- \sum_{kl} R(q)_{kl} \left[ \frac{\sum_{n} (\min_l r_{n|l}^q - r_{n|k}^q) }{\sum_{n} r_{n|k}^q}\right] \ln R(q)_{kl} \ \leq \ \tilde{{\mathcal{S}}}(R(q)) - {{\mathcal{S}}}(R(q)) \ \leq \ - \sum_{kl} R(q)_{kl} \left[ \frac{\sum_{n} (\max_l r_{n|l}^q - r_{n|k}^q) }{\sum_{n} r_{n|k}^q}\right] \ln R(q)_{kl} .
\end{eqnarray}
While the left inequality is always less than or equal to zero, the right inequality is always greater than or equal to zero. Clearly $\tilde{{\mathcal{S}}}(R(q)) = {{\mathcal{S}}}(R(q))$
when both inequalities are saturated at the same time. According to the Min-Max theorem~\cite{Hardy1959}, this happens if and only if all the $r_{n|l}$'s are equal (i.e., $l$ independent) for each fixed $n$. In other words, $\tilde{{\mathcal{S}}}(R(q)) = {{\mathcal{S}}}(R(q))$ if and only if $\tilde{R}(q)_{kl} = R(q)_{kl}$.
So again, we can track down our problem to the fact that $R(q)_{ij}$ is not a true joint distribution unless events are independent.

In case we wish to retain the form of the conditional entropy obtained from~(\ref{12a}) as defined in the J-A axioms, it is not difficult to find corrections to the $q$-additive entropic chain rule. In fact, from (\ref{13}) and (\ref{12aa}) we can see that we should employ the following substitution in the J-A additivity rule
\begin{eqnarray}
\mathcal{D}_q(B|A) \ \mapsto \ e^{\frac{(1-q)}{q} \left[\tilde{{\mathcal{S}}}(R(q)) - {{\mathcal{S}}}(R(q))\right]}\left(\mathcal{D}_q(B|A) + \frac{1}{1-q} \right) - \frac{1}{1-q} \ = \  \mathcal{D}_q(B|A)\  + \ {\mathcal{O}}(\tilde{{\mathcal{S}}}(R(q)) - {{\mathcal{S}}}(R(q)))   \, ,
\label{22aaab}
\end{eqnarray}
with ${\mathcal{O}}(\ldots)$ representing the error symbol.  Clearly the J-A additivity rule with the substitution (\ref{22aaab}) included reduces to the standard $q$-additivity form in the case when the two events are independent.

On the other hand, should we wish to retain the $q$-additive entropic chain rule together with the form of $\mathcal{D}_q$  defined in (\ref{hybrid})  we should
change the the definition of the conditional entropy according to prescription (\ref{13}).

%%%%%%%%%%%%%%%%%%%%%%%%%%%%%%%%%%%%%%%%%%%%%%%%%%%%%%%%%%%%%%%%%%%%%%%%
\section{Discussion}
%%%%%%%%%%%%%%%%%%%%%%%%%%%%%%%%%%%%%%%%%%%%%%%%%%%%%%%%%%%%%%%%%%%%%%%

In this Comment we have demonstrated that the hybrid entropies $\mathcal{D}_q$ introduced in~\cite{ja,jk} satisfy the Tsallis-type  $q$-additivity condition for any two independent events. As long as the events (or systems) considered are dependent, the $q$-additive entropic chain rule is not satisfied. We could trace down the root cause of this behavior to a peculiar trait of the  De~Finetti--Kolmogorov relation for escort distributions employed in the proof in Ref.~\cite{ja}. In particular, the latter is generally not implied by its non-escort counterpart.  This is notably reflected in the fact that $R(q)_{ij}$ is not a joint distribution although it is constructed directly from a genuine joint distribution $r_{ij}$. We have shown here that $R(q)_{ij}$ reduces to a joint distribution only for independent events, which in turn also defines the region of validity of the $q$-additive entropic chain rule for $\mathcal{D}_q$.

It should be noted, in passing, that in standard thermostatistical situations one deals only with entropies
containing independent subsystems. This is fully in accord with Landsberg's classification of thermodynamical systems with non-extensive entropies~\cite{Landsberg}.
In this respect, most of the results obtained for $\mathcal{D}_q$ in the literature can be used safely.

%%%%%%%%%%%%%%%%%%%%%%%%%%%%%%%%%%%%%%%%%%
\section*{Acknowledgments}
%%%%%%%%%%%%%%%%%%%%%%%%%%%%%%%%%%%%%%%%%%

%{\em {Acknowledgement.}} ---
We acknowledge the support by the GA\v{C}R Grant GA14-07983S.

%\vspace{-4mm}

%%%%%%%%%%%%%%%%%%%%%%%%%%%%%%%%%%%%%%%%%%%%%%%%%%%%%%%%%%%%%%%%%%%%%%%%
%\section*{References}
%%%%%%%%%%%%%%%%%%%%%%%%%%%%%%%%%%%%%%%%%%%%%%%%%%%%%%%%%%%%%%%%%%%%%%%

\end{document}